# *A Graph-Augmented knowledge Distillation based Dual-Stream Vision Transformer with Region-Aware Attention for Gastrointestinal Disease Classification with Explainable AI*

Md. Assaduzzaman[1]*, Nushrat Jahan Oyshi[1], Eram Mahamud[1]
[1] Department of Computer Science and Engineering
Daffodil International University
Dhaka 1207, Bangladesh

**Abstract**

The accurate and efficient classification of gastrointestinal (GI) diseases from endoscopic and histopathological imagery remains a significant challenge in medical diagnostics, mainly due to the vast data volume and subtle inter-class visual variations. This study presents a hybrid dual-stream deep learning framework built on teacher–student knowledge distillation, where a high-capacity teacher model integrates the global contextual reasoning of a Swin Transformer (Small) with the local fine-grained feature extraction of a Vision Transformer (ViT16 Small). The student network, implemented as a compact Tiny-ViT (5M parameters), inherits the teacher's semantic and morphological knowledge via a soft-label distillation process, achieving a balance between efficiency and diagnostic accuracy. Two carefully curated Wireless Capsule Endoscopy (WCE) datasets, encompassing four major GI disease classes normal mucosa, ulcerative colitis, polyps, and esophagitis, were employed to ensure balanced representation and prevent inter-sample bias. The proposed framework achieved remarkable classification performance, with accuracies of 99.78% and 99.28% on Dataset-1 and Dataset-2, respectively demonstrating strong discriminative performance. Interpretability analyses were conducted using Grad-CAM, Grad-CAM++, LIME, and Score-CAM to examine the regions contributing to the model's predictions. The resulting visual explanations indicated that the model primarily focused on tissue areas exhibiting relevant morphological characteristics. Furthermore, the Tiny-ViT student model achieved performance comparable to that of the transformer-based teacher while reducing computational complexity and inference time, suggesting its suitability for deployment in resource-constrained or real-time clinical settings. By balancing diagnostic rigor with computational efficiency, this distillation approach offers a practical, interpretable path for deploying advanced AI in routine endoscopic practice.

## 1. Introduction

Gastrointestinal (GI) diseases such as colorectal polyps, esophagitis, and ulcerative colitis are among the most common and potentially serious conditions affecting the human digestive system. If left undiagnosed, these diseases can lead to complications including chronic pain, malnutrition, and in some cases, colorectal cancer, one of the leading causes of cancer-related deaths globally. Early detection and timely treatment are essential for improving patient outcomes and reducing the global disease burden. One of the most effective diagnostic tools for identifying abnormalities in the GI tract is Wireless Capsule Endoscopy (WCE). WCE involves swallowing a small capsule equipped with a camera that captures thousands of high-resolution images as it passes through the digestive tract. While this technology has significantly improved diagnostic coverage, the sheer volume of images produced during a single session, often over 50,000 frames, creates a significant bottleneck for clinicians. Manually reviewing these images is time-consuming, labour-intensive, and prone to human error or oversight, particularly when abnormalities are subtle or infrequent [1]. In recent years, deep learning techniques, especially Convolutional Neural Networks (CNNs), have shown great potential in automating medical image classification tasks [2] Several studies have applied CNNs to endoscopy images and demonstrated promising results in detecting specific GI disorders. However, these methods are often limited by the quality and structure of the datasets used. Many public WCE datasets are either too small, unbalanced in class distribution, or contain overlapping samples across training and test splits, leading to biased evaluation and overfitting. In real clinical settings, such shortcomings can undermine the reliability and generalizability of AI models [3,4]. [suggestion: here, try to refer review paper on this domain published in your target journal]

To address the limitations, we utilized the WCE Curated Colon Disease Dataset, a well-structured and class-balanced dataset developed by aggregating and refining samples from established public sources, including KVASIR and ETIS-Larib-Polyp [5]. This curated dataset contains four clearly defined disease categories: normal mucosa, ulcerative colitis, polyps, and esophagitis with equal representation, ensuring more stable model training and fair evaluation. It was preprocessed and partitioned carefully to prevent data leakage between the training and testing sets, thereby preserving the validity of the reported performance metrics [6]

Motivated by the identified research gap, we propose a deep learning framework for automated classification of gastrointestinal (GI) conditions from endoscopic images. The model is designed to capture both local and global visual representations through multi-scale feature extraction and attention mechanisms. Additionally, preprocessing steps, including image enhancement and data augmentation, are incorporated to improve robustness and generalization to unseen samples. The overall training pipeline is structured to mitigate overfitting while maintaining consistent classification performance across the four disease categories.

The main contributions of this paper are as follows:

• We propose a dual-stream transformer-based framework that integrates Swin Transformer and ViT-16 architectures to capture complementary global contextual and fine-grained morphological features for gastrointestinal disease classification.

• A lightweight Tiny-ViT student model is developed using knowledge distillation, enabling efficient inference while maintaining competitive diagnostic performance across wireless capsule endoscopy and colorectal histopathology datasets.

• Region-aware attention, multi-scale feature fusion, and multiple explainable AI techniques are incorporated to enhance feature representation and evaluate the interpretability of model predictions.

## 2. Related Works

Artificial Intelligence (AI) has significantly advanced medical diagnostics, particularly in the detection and classification of gastrointestinal (GI) and colorectal cancers (CRC). Deep learning architectures, including convolutional neural networks (CNNs) and transformer-based models, have enabled automated analysis of large volumes of endoscopic and histopathological data. These approaches have improved diagnostic accuracy, reduced manual workload, and enhanced the interpretability of medical imaging systems.

Deep Learning for Histopathological Image Classification

Several studies have focused on applying deep learning techniques to classify colorectal cancer tissues from histopathological images. Le et al. [7] proposed a CNN-based framework that achieved high classification performance by focusing on fine-grained morphological features of colorectal tissues. Similarly, Ömeroğlu [8] introduced a supervised contrastive learning approach to enhance feature discrimination between tumour and non-tumour regions, improving representation learning in CRC datasets. Fadafen and Rezaee [9] further extended this direction by proposing an ensemble-based hybrid deep learning framework for multi-tissue CRC classification, demonstrating improved robustness across heterogeneous datasets. Transformer-based architectures have also been explored to capture long-range contextual relationships in medical images. El Amine et al. [10] proposed a semi-supervised transformer network incorporating Generative Adversarial

Networks (GANs) for image normalisation and attention mechanisms to identify diagnostically relevant regions. Their framework utilised Vision Transformer–based knowledge distillation to improve generalisation. Likewise, Ke et al. [11] developed a multi-magnification ensemble combining transfer learning and CNN architectures, achieving strong performance across different magnification scales. While these approaches demonstrate high accuracy, many of them rely on carefully curated datasets and may struggle to generalise across institutions due to variations in staining protocols and imaging conditions. Nevertheless, these studies highlight the effectiveness of deep learning models in extracting discriminative histopathological features for CRC detection.

AI-Based Analysis of Endoscopic and Capsule Imaging

Beyond histopathology, AI has been widely applied to endoscopic imaging modalities, particularly Wireless Capsule Endoscopy (WCE), which produces thousands of frames during a single examination. Mirza et al. [12] developed a fused architecture integrating shallow and deep neural networks to classify WCE images into multiple disease categories. Bordbar et al. [13] proposed a 3D convolutional network that captures spatial–temporal relationships across consecutive frames, improving the detection of abnormalities such as polyps and ulcers. Interpretability and reliability in clinical decision-making have also received increasing attention. Shukla and Jayaraman [14] incorporated explainable AI (XAI) techniques, including Grad-CAM and LIME, to visualise regions influencing abnormality detection in endoscopic images. Attallah et al. [15] proposed EndoNet, a multi-scale attention-based network designed to enhance disease detection across multiple endoscopic datasets. Similarly, Tan et al. [16] introduced EndoOOD, an uncertainty-aware out-of-distribution detection framework that improves robustness when models encounter unseen data in real-world deployments. Although these methods significantly improve automated diagnosis, many existing systems still analyse individual frames independently and often fail to exploit the full temporal context present in WCE video streams.

Lightweight and Interpretable Diagnostic Models

In addition to high-performance architectures, several researchers have emphasised computational efficiency and model interpretability for real-world clinical deployment. Ömeroğlu [17] designed a compact transfer learning–based model for CRC tissue recognition that reduces computational requirements while maintaining competitive accuracy. Alotaibi et al. [18] proposed HIELCC-EDL, a hybrid system integrating Wiener filtering with a swarm-optimised ensemble of classifiers, achieving an accuracy of 99.6% on histopathology images. Explainability has also been integrated into diagnostic frameworks. Alsubai et al. [29] combined local binary pattern (LBP) features with the Inception-ResNetV2 architecture and incorporated SHapley Additive exPlanations (SHAP) to improve model interpretability. While these approaches provide valuable insights into model decision-making, many explainability methods still offer coarse visual explanations that may not fully correspond to clinically meaningful anatomical structures.

Multi-Modal and Biomarker-Based AI Approaches

Recent research has also explored the integration of multi-modal data and biological biomarkers to enhance diagnostic and prognostic analysis. Huang et al. [20] proposed an AI-based survival stratification system for stage II CRC patients by combining CT imaging features with pathological markers. Tafavvoghi et al. [27] developed a history-based deep learning framework capable of identifying mismatch repair deficiency in CRC tumour slides. Westwood et al. [28] analysed tumour-infiltrating lymphocytes (TILs) and tumour cell density (TCD) from whole-slide images, demonstrating their prognostic value for CRC risk prediction. Advancements in imaging technologies have further enabled nanoscale biomarker analysis. Studies using chromatin-sensitive partial wave spectroscopy (csPWS) have revealed chromatin structural alterations associated with colorectal cancer risk [30]. Similarly, multi-stain deep learning models have been developed to integrate diverse staining patterns and immune cell distributions, improving model generalisation and biological interpretability [34]. Although these approaches provide valuable clinical insights, they often require complex multi-modal data acquisition and specialised imaging protocols that may limit their scalability in routine clinical workflows.

Despite the substantial progress achieved in AI-driven CRC diagnostics, several limitations remain. Many existing frameworks rely on single-modality datasets and struggle with cross-centre generalisation. Additionally, most studies focus primarily on classification performance while providing limited interpretability and failing to integrate both global contextual information and fine-grained morphological features within a unified architecture. To address these challenges, the present study proposes a dual-stream hybrid deep learning framework that integrates global contextual representations from a Swin Transformer Small with local feature extraction from a ViT-16 Small model. A teacher–student knowledge distillation strategy is employed to transfer rich semantic and morphological knowledge from the transformer teacher network to a lightweight Tiny-ViT student model, enabling high diagnostic accuracy, improved interpretability, and computational efficiency suitable for real-time clinical applications.

Table 1. Critical Analysis of Related Works on AI-Based Colorectal Cancer (CRC) Detection

| **Author / Year** | **Focus Area** | **Model / Technique** | **Key Limitation Relevant to This Study** |
|---|---|---|---|
| Le et al., 2025 [7] | CRC histopathology classification | CNN-based hierarchical classifier | CNN captures local features but struggles to model long-range contextual relationships in tissue structures |
| Ömeroğlu, 2024 [8] | CRC histopathology via contrastive learning | Supervised contrastive learning with ResNet backbone | Representation learning improves discrimination but still relies on CNN features with limited global context modelling |
| Fadafen & Rezaee, 2023 [9] | Multi-tissue CRC classification | Ensemble hybrid deep learning framework | Ensemble improves accuracy but introduces high computational complexity and deployment cost |
| El Amine et al., 2023 [10] | Semi-supervised CRC classification | Vision Transformer + GAN normalisation + knowledge distillation | Complex multi-stage pipeline that is computationally expensive for real-time clinical use |
| Ke et al., 2025 [11] | Multi-magnification CRC classification | CNN ensemble (ResNet, DenseNet) | Magnification-specific models may overfit and lack unified feature representation across scales |
| Mirza et al., 2025 [12] | GI disease classification (WCE) | Hybrid shallow–deep CNN fusion | Frame-level analysis without capturing broader contextual relationships |
| Bordbar et al., 2023 [13] | Capsule endoscopy video analysis | 3D CNN architecture | High memory and computational requirements for large-scale clinical deployment |
| Attallah et al., 2025 [15] | Endoscopic disease detection | EndoNet (multi-scale CNN with attention) | High model complexity and limited validation across heterogeneous datasets |
| Tan et al., 2024 [16] | Out-of-distribution detection in endoscopy | Transformer-based uncertainty modelling | Focused primarily on robustness rather than improving diagnostic representation learning |
| Author / Year | Focus Area | Model / Technique | Key Limitation Relevant to This Study |

| Le et al., 2025 [7] | CRC histopathology classification | CNN-based hierarchical classifier | CNN captures local features but struggles to model long-range contextual relationships in tissue structures |
|---|---|---|---|
| Ömeroğlu, 2024 [8] | CRC histopathology via contrastive learning | Supervised contrastive learning with ResNet backbone | Representation learning improves discrimination but still relies on CNN features with limited global context modelling |
| Fadafen & Rezaee, 2023 [9] | Multi-tissue CRC classification | Ensemble hybrid deep learning framework | Ensemble improves accuracy but introduces high computational complexity and deployment cost |
| El Amine et al., 2023 [10] | Semi-supervised CRC classification | Vision Transformer + GAN normalisation + knowledge distillation | Complex multi-stage pipeline that is computationally expensive for real-time clinical use |
| Ke et al., 2025 [11] | Multi-magnification CRC classification | CNN ensemble (ResNet, DenseNet) | Magnification-specific models may overfit and lack unified feature representation across scales |

Despite these advances, several challenges remain in AI-based colorectal cancer diagnosis. Many existing approaches rely on convolutional architectures that primarily capture local morphological features while lacking the ability to effectively model global contextual relationships within histopathological structures. Conversely, transformer-based models provide stronger global representation learning but often require large computational resources and complex training pipelines, which can limit their practicality in real-world clinical settings. Furthermore, ensemble frameworks designed to improve performance frequently increase model complexity and inference latency, making them less suitable for efficient deployment.To address these limitations, this study proposes a dual-stream hybrid architecture that integrates global contextual representations from a Swin Transformer Small model with local feature extraction from a ViT-16 Small model. Through a teacher–student knowledge distillation framework, the proposed approach transfers both semantic and morphological knowledge to a lightweight Tiny-ViT student model, enabling efficient and accurate colorectal cancer classification while maintaining computational feasibility for clinical applications.

## 3. Methodology

### 3.1. Overview

The proposed framework introduces a dual-encoder knowledge distillation paradigm for the automated classification of gastrointestinal (GI) diseases from Wireless Capsule Endoscopy (WCE) images. The architecture adopts a teacher–student learning strategy, where the teacher network integrates two complementary vision transformers, Swin Transformer Small (patch4-window7-224) as the global encoder and ViT-Small (patch16-224) as the local encoder. This combination enables the extraction of both macro-level contextual features and micro-level texture representations, essential for differentiating subtle pathological variations in endoscopic imagery. The Swin Transformer captures hierarchical and spatial dependencies across vast receptive fields, while the ViT16 component focuses on localised morphological cues such as glandular structures and mucosal irregularities.

The student network, implemented using Tiny-ViT (tiny_vit_5m_224), provides a lightweight, computationally efficient alternative that can reproduce the teacher's diagnostic capability under constrained resources. [35,36] Through knowledge distillation, the student model assimilates both the explicit supervision from ground-truth labels and the implicit knowledge contained in the teacher's soft probability distributions, thereby inheriting its discriminative and interpretive strengths while maintaining rapid inference. The overall methodological pipeline, illustrated schematically in Figure 1, encompasses dataset curation, image preprocessing and augmentation, dual-stream architectural design, knowledge transfer via distillation, and comprehensive performance evaluation across multiple metrics and interpretability frameworks. This design ensures a robust, interpretable, and scalable solution for endoscopic disease classification suitable for real-time clinical decision support.

## 3.2. Dataset Description

### 3.2.1 Dataset 1

Dataset 1 comprises endoscopic imagery sourced from two publicly available, clinically curated repositories, primarily the Kvasir dataset [37]. These images, captured via Wireless Capsule Endoscopy (WCE), were released following standardized preprocessing and clear source separation to mitigate the risk of patient-level data leakage across training and validation splits. Adopting the distribution strategy established in prior literature (e.g., Montalbo, 2022), we maintained the original folder structure and partitioning to ensure benchmarking consistency. Data preparation was limited to essential preprocessing for model ingestion, specifically bilinear resizing to the network's input dimensions and per-channel normalization. The final curated set targets four diagnostic categories representative of common WCE findings: *normal mucosa, ulcerative colitis, colorectal polyps, and esophagitis*. Ground-truth labels were inherited from the source repositories and mapped to the standardized class identifiers used throughout this study.

Dataset distribution

The Kaggle release contains 6,000 WCE images organised into train/, val/, and test/ partitions and four class folders per split (0_normal, 1_ulcerative_colitis, 2_polyps, 3_esophagitis). The test partition is explicitly balanced with 200 images per class (total 800 test images). The train/ and val/ partitions follow the same four-class structure and are used as released; class balance is preserved per split for fair evaluation. Table 1 reports the composition of the test split (the portion with verified per-class counts), which we use for final model assessment. Counts for training and validation are taken from the released split and are described alongside implementation details in Section 4 (*Experiments*). Table 2 shows the composition of dataset 1 composition.

Table 2. Dataset 1 (Kaggle WCE Curated) test split composition

| Dataset 1 | | |
|---|---|---|
| **Class Label** | **Pathology Type** | **Test** |
| 0_normal | Normal mucosa | 200 |
| 1_ulcerative_colitis | Ulcerative colitis | 200 |
| 2_polyps | Colorectal polyps | 200 |
| 3_esophagitis | Esophagitis | 200 |

### 3.2.2 Dataset 2

Dataset 2 comprises histopathological image patches derived from hematoxylin and eosin (H&E) stained tissue samples of colorectal cancer, collected from 17 patients in the Central Finland Healthcare District. The dataset is publicly available via Mendeley [38]. The dataset contains 2,770 image patches of 224×224 pixels at 0.5 μm per pixel, distributed into three tissue classes: tumour, stroma, and other (including debris, lymphocytes, mucus, normal epithelium, and smooth muscle). Labels were acquired directly from pathologist annotations, which are mapped to consistent class identifiers used throughout the dataset. **Table 3** shows the patch distribution per class.

Table 3. Dataset 2 (Mendeley Colorectal Histopathology Patches)-Class Distribution

| Class | Number of Patches |
|---|---|
| Tumor | 1,004 |
| Stroma | 1,001 |
| Other | 765 |
| Total | 2,770 |

**Dataset distribution**

Depending on the experimental needs, the train/validation/test were distributed, maintaining a 75:15:15 ratio. The dataset is intended for tasks such as external testing of deep learning models for tissue classification and segmentation, as well as for digital pathology studies. The training, validation, and test partitions are balanced across all categories. The distribution is shown in Table 4.

Table 4. Dataset-2 composition and distribution

| Class Label | Pathology Type | Train | Validation | Test |
|---|---|---|---|---|
| 0_tumor | Tumour | 754 | 125 | 125 |
| 1_stroma | Stroma | 751 | 125 | 125 |
| 2_other | Other | 575 | 95 | 95 |
| **Total** | — | 2,080 | 345 | 345 |

### 3.3. Preprocessing and Augmentation

To standardize the input format and ensure compatibility with pretrained architectures, all images were resized to 224×224 pixels. Pixel intensities were normalized on a per-channel basis using the ImageNet statistics: mean $\mu = [0.485, 0.456, 0.406]$ and standard deviation $\sigma = [0.229, 0.224, 0.225]$, thereby aligning the dataset's pixel distribution with the pretraining domain. For the training subset, a constrained augmentation pipeline was applied to improve generalization while preserving the integrity of pathological features. This included random horizontal flipping to simulate natural variations in capsule orientation, small in-plane rotations (±10°) to reflect realistic endoscopic motion, and mild affine transformations to mimic variations in viewpoint and position. These augmentations were designed to expand the representational diversity of the training data without introducing artifacts that could compromise clinical interpretability. The validation and test subsets underwent only resizing and normalization to maintain a consistent and unbiased evaluation protocol.

3.4. Model Architectures

The overall workflow of the proposed framework is illustrated in **Figures 1 and 2, which summarise** the end-to-end architecture, including data flow, dual-encoder teacher network, knowledge distillation pipeline, and the lightweight Tiny-ViT student model.

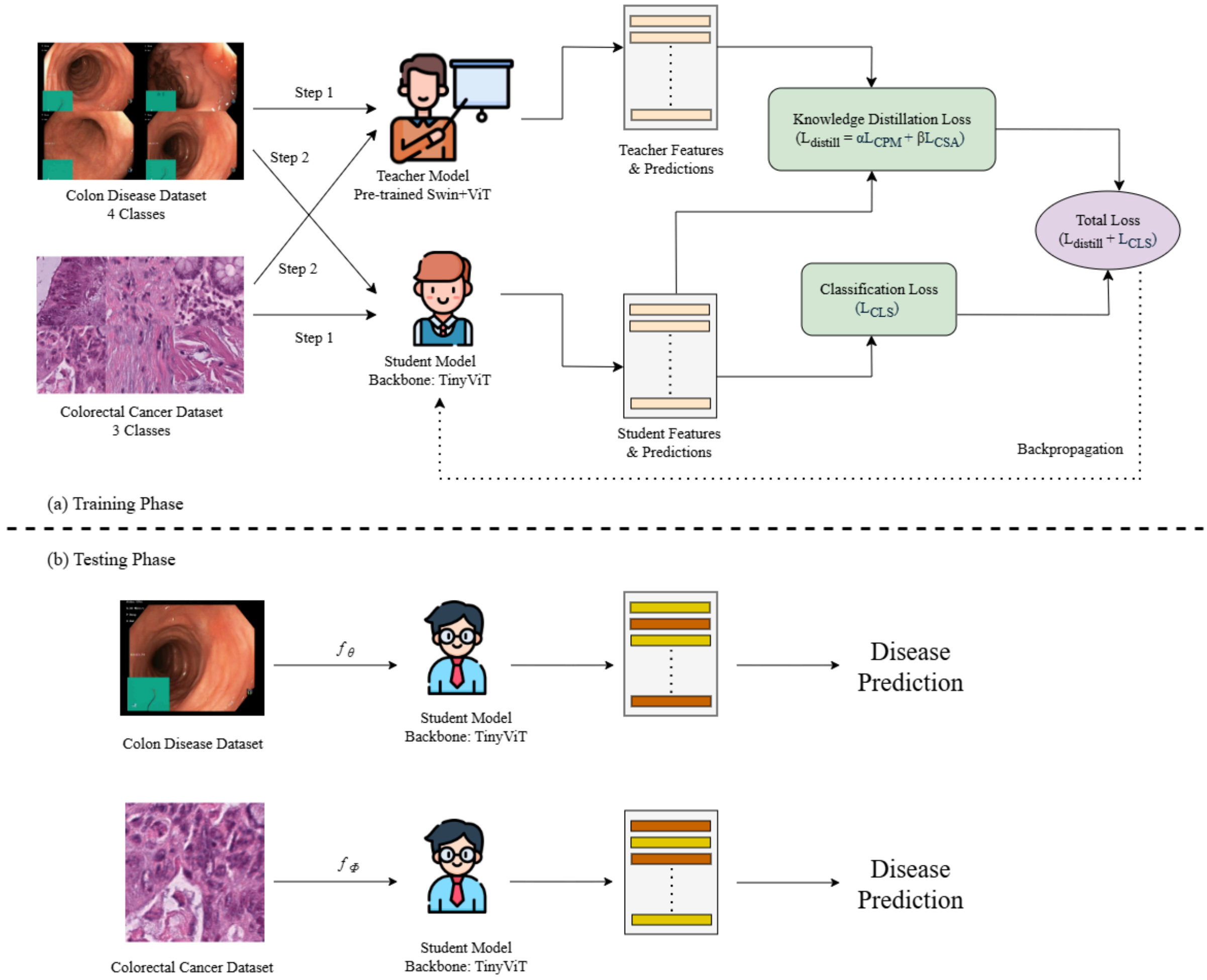

Fig 1. Training and testing phases of the proposed teacher–student knowledge distillation framework.

The proposed framework adopts a teacher–student architecture that integrates a dual-encoder Transformer teacher with a lightweight Tiny-ViT student, designed to achieve both high diagnostic accuracy and computational efficiency in gastrointestinal (GI) disease classification from Wireless Capsule Endoscopy (WCE) imagery. The teacher network incorporates two complementary encoders: a Swin Transformer (Small, patch4-window7-224) serving as the global encoder, and a Vision Transformer (ViT-Small, patch16-224) acting as the local encoder. The Swin Transformer employs a shifted-window self-attention mechanism that hierarchically models long-range dependencies across spatial regions with linear computational complexity relative to image size. This enables the network to capture large-scale contextual relationships such as mucosal patterns and diffuse inflammation.

In contrast, the ViT16 module focuses on fine-grained local textures and micro-structural cues, including glandular boundaries, epithelial distortions, and lesion morphology. Features extracted from both encoders are fused using an attention-guided feature aggregation module, which adaptively balances global and local representations. The dual-encoder design thus integrates macroscopic contextual awareness with microscopic precision, providing a comprehensive representation of GI pathology. Both encoders are initialised with ImageNet-1K weights, and early layers are partially frozen to preserve low-level feature generalisation while deeper layers are fine-tuned for domain-specific adaptation. The student model is implemented as Tiny-ViT (tiny_vit_5m_224), a compact Vision Transformer containing approximately 5 million parameters [39]. It retains the essential self-attention structure of standard ViTs while employing depth-reduced transformer blocks and efficient patch embeddings to achieve low latency and reduced memory consumption.

During the distillation process, the Tiny-ViT student learns from both hard labels and soft targets generated by the dual-encoder teacher, effectively inheriting its decision boundaries and inter-class relationships. [40] The final classification head outputs four softmax-activated probabilities corresponding to the diagnostic categories: normal mucosa, ulcerative colitis, polyps, and esophagitis. This hybrid configuration ensures that the teacher network provides rich, hierarchical supervision. At the same time, the student model delivers lightweight, real-time inference, a crucial balance for practical deployment in clinical endoscopy workflows.

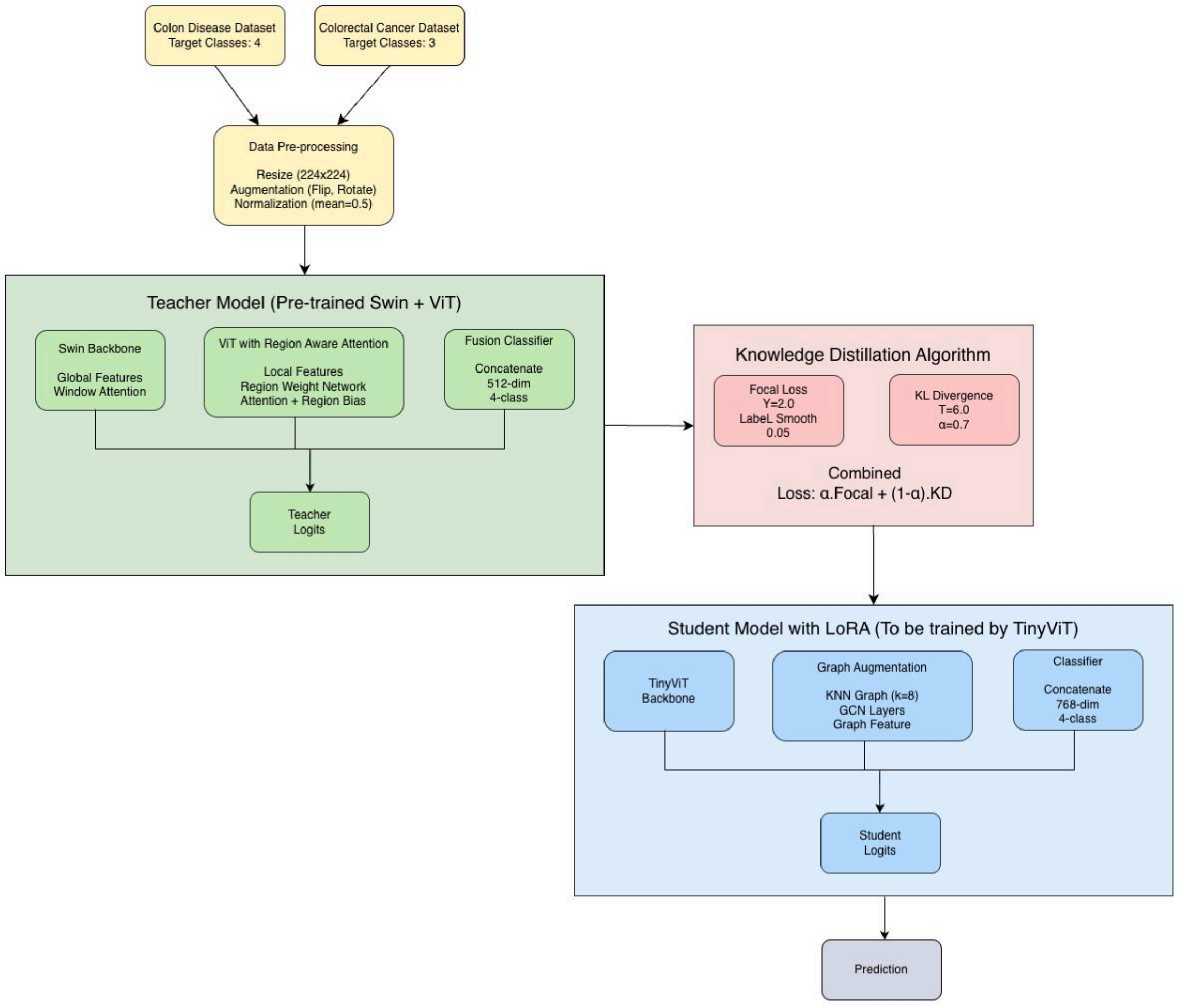

Fig 2. Detailed architectural components of the teacher model and LoRA-based Tiny-ViT student model.

### 3.4.1. Teacher Model: Swin Transformer + ViT16

The teacher network integrates a dual-encoder architecture that combines a global encoder (Swin Transformer Small, patch4-window7-224) with a local encoder (ViT-Small, patch16-224) to capture complementary global and fine-grained contextual information from endoscopic imagery. [41] The Swin Transformer component employs a shifted-window attention mechanism, enabling efficient modelling of hierarchical spatial dependencies with linear computational complexity relative to image size, as shown in Fig. 3. This allows the model to preserve structural integrity while effectively capturing large-scale contextual variations such as mucosal pattern shifts or diffuse inflammation. In contrast, the ViT16 module processes finer local image patches, focusing on micro-textural features, such as glandular boundaries, epithelial disruptions, and localised lesions, which are critical for GI disease diagnosis. Feature representations from both encoders are fused through an attention-based feature fusion layer, which adaptively weights the contribution of global and local representations before classification. Parameters were initialised using ImageNet pretraining, with early layers frozen to preserve generic visual features, while higher transformer blocks were fine-tuned for domain-specific adaptation to gastrointestinal pathology. This dual-encoder design enables the teacher model to serve as a powerful feature extractor, integrating regional precision and global contextual awareness.

Window partitioning

$$X_w = Partition(X) \quad (2)$$

Self-Attention inside each window

$$Attention(Q, K, V) = Softmax(\frac{QK^T}{\sqrt{d_k}} + B)V \quad (3)$$

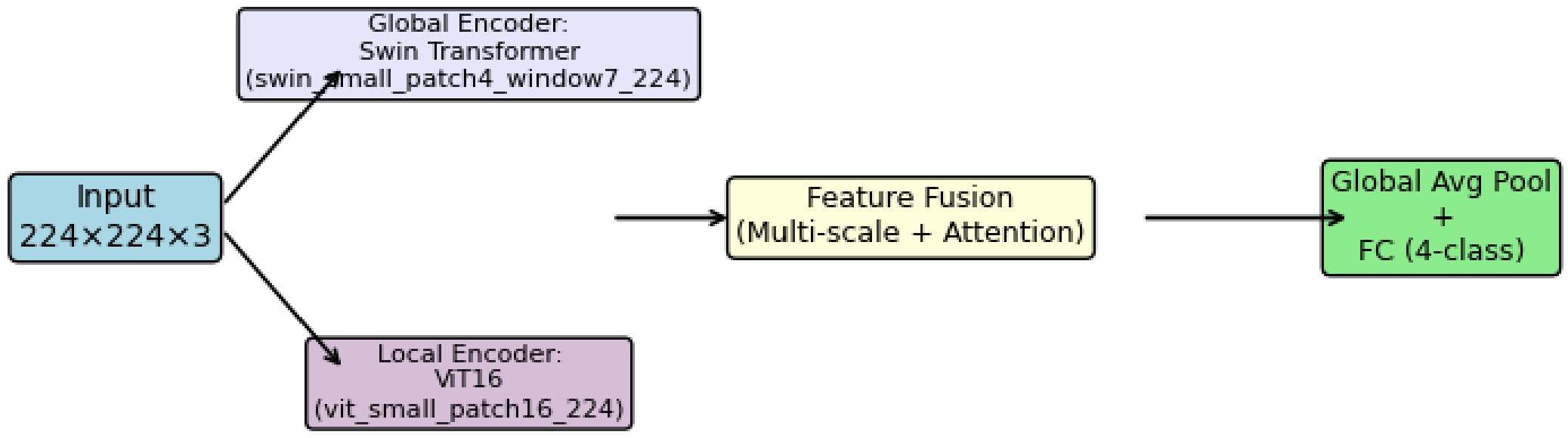

Fig 3. Teacher Model- Swin Transformer + ViT16

3.4.2. Student Model: Tiny-ViT (tiny_vit_5m_224)

The student network employs a Tiny Vision Transformer (Tiny-ViT, 5M parameters), a lightweight, computationally efficient architecture that replicates the diagnostic performance of the teacher while maintaining high inference speed and deployability. In Fig 4, Tiny-ViT utilises optimised patch embeddings and compact attention blocks that preserve long-range spatial relationships while significantly reducing parameter overhead. During training, the student model learns from both hard labels (ground-truth classes) and soft labels produced by the teacher network through knowledge distillation. The soft targets convey inter-class relationships and confidence distributions, allowing the student to internalise the teacher's learned decision boundaries. Pretrained on ImageNet and fine-tuned for the four-class WCE dataset, the Tiny-ViT model demonstrates excellent balance between accuracy and efficiency. [42] Its reduced memory footprint and inference time make it particularly well-suited for real-time computer-aided diagnosis (CAD) and edge-device deployment in clinical endoscopy systems.

ViT Patch Embedding

$$z_0 = [x_p^1 E; x_p^2 E; \ldots ; x_p^N E] + E_{pos} \quad (4)$$

Where
– $E$patch embedding projection
– $E_{pos}$ = positional encoding

Multi-Head Self-Attention (MHSA)

$$MHSA(Q, K, V) = Concat(h_1, h_2, ..., h_H)W^O \quad (5)$$

$$h_i = Softmax(\frac{QW_i^Q(KW_i^K)^T}{\sqrt{d_k}})VW_i^V \quad (6)$$

Feed Forward Network

$$FFN(x) = GELU\left(xW_1 + b_1\right)W_2 + b_2 \quad (7)$$

Region-Aware Attention (Local Encoder)

$$A = Softmax(\frac{QK^T}{\sqrt{d_k}} + R) \quad (8)$$

Where
$CapR$= region bias from region weighting network.

Attention-Guided Feature Fusion

Let $F_g$ = global features (Swin)
Let $F_l$ = local features (ViT)

Fusion attention weight

$$\alpha = \sigma(W[F_g \parallel F_l]) \quad (9)$$

Final fused feature

$$F = \alpha F_g + (1 - \alpha)F_l \quad (10)$$

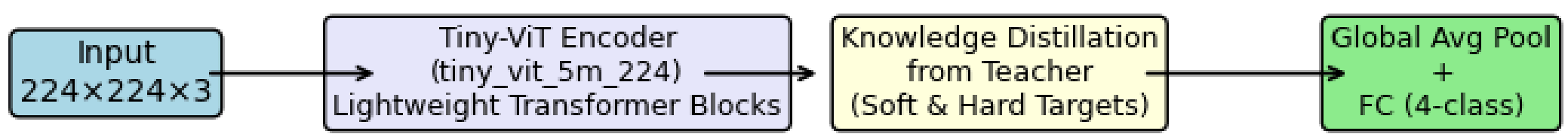

Fig 4. Student Model- Tiny-ViT

3.5. Knowledge Distillation Framework

The KD strategy combines **hard-target supervision** using cross-entropy loss with **soft-target supervision** via Kullback–Leibler (KL) divergence between temperature-scaled softmax outputs of the teacher and student.

The final loss is formulated as:

$$Ltotal = \alpha T2KL(\sigma(zT/T)||\sigma(zS/T)) + (1 - \alpha)LCE(y, \sigma(zS)) \quad (11)$$

Where:

- $zT$ and $zS$ are teacher and student logits,
- $T = 4. is$ the temperature scaling factor,

- $\alpha = 0.9$ balances the soft and hard loss components,
- $\sigma(\cdot$ is the softmax function.

**3.6. Training protocol**

Both the teacher and student networks were trained under identical optimisation conditions to ensure stable and comparable convergence. The models were trained using the AdamW optimiser with an initial learning rate of $1 \times 10^{-4}$, weight-decay regularisation, and a batch size of 32. [56] A ReduceLROnPlateau scheduler with a reduction factor of 0.1 and a patience of three epochs was employed to automatically decrease the learning rate when the validation loss plateaued. To avoid overfitting, early stopping was implemented with a patience of five epochs, preserving the best model weights based on the lowest validation loss achieved during training. All models were trained for a maximum of 50 epochs, and gradient clipping was applied to stabilise updates during the early training phase. For the teacher model, a hybrid dual-encoder architecture was utilised, combining a global encoder (Swin Transformer – swin_small_patch4_window7_224) and a local encoder (ViT – vit_small_patch16_224). Their feature representations were fused through multi-scale attention before classification. The student model, implemented as Tiny-ViT (tiny_vit_5m_224), was trained under a knowledge-distillation regime, learning both from ground-truth labels (hard targets) and the soft logits of the teacher network. The total loss combined the cross-entropy loss and distillation loss, balanced by coefficients $\alpha = 0.9$ and temperature $T = 4$. The complete training configurations are summarised in Table 5.

Table 5. Training configuration and hyperparameters

| **Parameter** | **Teacher Model (Swin Transformer + ViT16)** | **Student Model (Tiny-ViT 5M)** |
|---|---|---|
| Global Encoder | Swin Transformer (swin_small_patch4_window7_224) | – |
| Local Encoder | ViT (small_patch16_224) | – |
| Optimizer | AdamW | AdamW |
| Learning Rate | $1 \times 10^{-4}$ | $1 \times 10^{-4}$ |
| Batch Size | 32 | 32 |
| Epochs (Max) | 50 | 50 |
| Scheduler | ReduceLROnPlateau (factor = 0.1, patience = 3) | ReduceLROnPlateau (factor = 0.1, patience = 3) |
| Early Stopping Patience | 5 | 5 |
| Weight Decay | Applied | Applied |
| Loss Function | Cross-Entropy | CE + KD ($\alpha = 0.9$, $T = 4$) |
| Feature Fusion | Multi-scale + Attention | – |

3.7. Evaluation metrics

The models were evaluated on a held-out test set using a combination of class-wise and aggregate performance metrics [57]. Overall classification accuracy was calculated together with precision, recall, and F1-scores for each class. Macro- and weighted-average scores were also reported to account for potential class imbalance. To analyse class-specific prediction behaviour, a confusion matrix was used to visualise misclassification patterns. In addition, a one-vs-rest receiver operating characteristic (ROC) analysis was performed for each class, where the area under the ROC curve (AUC) was used as a quantitative measure of class separability. The experimental results demonstrated very high classification performance. The distilled student model achieved an overall accuracy of **99.78%**, indicating that the knowledge distillation framework effectively preserved the discriminative capacity of the teacher network within a more compact architecture. The ROC analysis yielded an **AUC value close to 1.0000 across all classes**, suggesting strong separability between categories in the learned feature space. Although perfect or near-perfect AUC values can sometimes indicate potential overfitting, several factors support the reliability of this result. First, evaluation was conducted on a **strictly separated test** **set that was not used during training or hyperparameter tuning**, reducing the risk of information leakage. Second, the dataset contains **well-defined pathological patterns that are visually distinctive across classes**, which can lead to highly separable representations when deep transformer-based models are trained effectively. Finally, the **teacher–student knowledge distillation framework encourages smoother decision boundaries and improved generalisation**, which may contribute to the observed high separability. Nevertheless, future work will involve **cross-dataset validation and multi-centre evaluation** to further verify the robustness of the model in diverse clinical settings.

AdamW Update

$$m_t = \beta_1 m_{t-1} + (1 - \beta_1) g_t v_t = \beta_2 v_{t-1} + (1 - \beta_2) g_t^2 \hat{m}_t : \quad (12)$$

Weight decay

$$w_{t+1} = w_t - \eta(\hat{m}_t / \sqrt{\hat{v}_t} + \lambda w_t) \quad (13)$$

Evaluation Metrics Equations

Accuracy

$$Accuracy = \frac{TP+TN}{TP+TN+FP+FN} \quad (14)$$

Precision

$$Precision = \frac{TP}{TP+FP} \quad (15)$$

Recall

$$Recall = \frac{TP}{TP+FN} \quad (16)$$

F1-Score

$$F1 = 2\times \frac{Precision \cdot Recall}{Precision+Recall} \quad (17)$$

Macro Average

$$Macro - F1 = \frac{1}{K} \sum_{k=1}^{K} F1_k \quad (18)$$

Weighted Average

$$Weighted - F1 = \sum_{k=1}^{K} w_k F1_k \quad (19)$$

Where

$$w_k = \frac{support_k}{N}. \quad (20)$$

Confusion Matrix Equation

$$C_{ij} = |\{x: y = i, \hat{y} = j\}| \tag{21}$$

Where

$i$ = actual class,

$j$ = predicted class.

ROC & AUC Equations

True Positive Rate

$$TPR = \frac{TP}{TP+FN} \tag{22}$$

False Positive Rate

$$FPR = \frac{FP}{FP+TN} \tag{23}$$

AUC

$$AUC = \int_{0}^{1} TPR(FPR)\, d(FPR) \tag{24}$$

Grad-CAM

$$\alpha_k = \frac{1}{Z}\sum_i \sum_j \frac{\partial y^c}{\partial A^k_{ij}} L^c_{GradCAM} = ReLU(\sum_k \alpha_k A^k) \tag{25}$$

Grad-CAM++

$$: The\alpha^k_{ij} = \frac{\partial^2 y^c / \partial (A^k_{ij})^2}{2\frac{\partial^2 y^c}{\partial (A^k_{ij})^2} + \sum_a \sum_b A^k_{ab} \frac{\partial^3 y^c}{\partial (A^k_{ab})^3}} L^c_{GradCAM++} = ReLU(\sum_k \alpha^+_k A^k) \tag{26}$$

## 4. Results and Analysis

This section presents a comprehensive evaluation of the proposed teacher-student knowledge distillation framework. The models were assessed on the held-out test set using a suite of quantitative metrics, and the student model's decision-making process was further interrogated through qualitative visualization.

### 4.1 Quantitative Performance and Knowledge Distillation

Across two independent datasets, knowledge distillation produced a compact student model (~5.07M parameters) that closely approximated the performance of the high-capacity teacher while maintaining strong fit on the training and validation splits, as summarized in Table 6. On Dataset-2, the teacher achieved accuracies of 100.00%, 99.52%, and 98.80% on the training, validation, and test sets, respectively, with corresponding losses of 0.0189, 0.0262, and 0.0304. The distilled student attained accuracies of 100.00%, 99.28%, and 99.28% with losses of 0.0187, 0.0307, and 0.0323, respectively. Class-wise evaluation on Dataset-2 confirmed balanced performance across all classes (overall accuracy = 0.9928 over 417 images), including a Tumour precision of 1.0000, Stroma recall of 1.0000, and a macro-F1 score of 0.9927.

On Dataset-1, the teacher achieved 100.00%, 99.67%, and 99.67% accuracy (losses: 0.0304, 0.0379, 0.0373), while the student reached 99.98%, 99.67%, and 99.78% (losses: 0.0307, 0.0352, 0.0368). These results indicate that knowledge distillation acts as an effective regularizer, producing a smaller model that preserves in-distribution performance while maintaining or improving generalization on unseen data.

Table 6. Quantitative Performance and Knowledge Distillation

| **Dataset** | **Data Split** | **Metric** | **Teacher Model** | **Student Model** |
|---|---|---|---|---|
| **Dataset-2** | Train | Loss | 0.0189 | 0.0187 |
| | | Accuracy | 100.00% | 100.00% |
| | Validation | Loss | 0.0262 | 0.0307 |
| | | Accuracy | 99.52% | 99.28% |
| | Test | Loss | 0.0304 | 0.0323 |
| **Dataset-1** | | Accuracy | 98.80% | 99.28% |
| | Train | Loss | 0.0304 | 0.0307 |
| | | Accuracy | 100.00% | 99.98% |
| | Validation | Loss | 0.0379 | 0.0352 |
| | | Accuracy | 99.67% | 99.67% |
| | Test | Loss | 0.0373 | 0.0368 |
| | | Accuracy | 99.67% | 99.78% |

4.2 Class-Specific Classification Performance

On Dataset-2, evaluated over the 417-image test set (Table 7), the distilled student achieved an overall accuracy of 99.28% with consistently strong performance across classes. Stroma was fully identified (recall = 1.0000; 146/146 correct), while Tumour predictions were conservative, with precision = 1.0000 (no false positives) and a near-perfect recall of 0.9852 (two Tumour cases classified as Other). The Other category remained robust (precision = 0.9854; recall = 0.9926), with a single sample misclassified as Stroma. These results yielded high and balanced summary scores (macro-F1 = 0.9927; weighted-F1 = 0.9928), indicating reliable performance across all classes rather than improvements concentrated in a single category.

For Dataset-1, the distilled student also demonstrated improved deployment performance, achieving 99.78% test accuracy (loss = 0.0368) compared with the teacher's 99.67% (loss = 0.0373), while maintaining near-ceiling training accuracy (99.98%, loss = 0.0307) and comparable validation results (99.67%, loss = 0.0352) [45]. Although per-class precision, recall, F1 scores, and confusion matrices were not computed for this run, the close alignment of validation and test accuracies, together with the student's consistent test-set improvement over the teacher, supports balanced generalization rather than class-specific overfitting [46].

Table 7. Class-Specific Classification Performance

| **Dataset** | **Method** | **Class** | **Precision** | **Recall** | **F1-Score** | **Support** |
|---|---|---|---|---|---|---|
| Dataset-1 | Student Model with LoRA trained by TinyViT | 0_normal | 1.0000 | 1.0000 | 1.0000 | 225 |
| | | 1_ulcerative_colitis | 0.9912 | 1.0000 | 0.9956 | 225 |
| | | 2_polyps | 1.0000 | 0.9911 | 0.9955 | 225 |
| | | 3_esophagitis | 1.0000 | 1.0000 | 1.0000 | 225 |
| | | Accuracy | | | 0.9978 | 900 |
| | | Macro Avg | 0.9978 | 0.9978 | 0.9978 | 900 |
| | | Weighted Avg | 0.9978 | 0.9978 | 0.9978 | 900 |
| Dataset-2 | | Other | 0.9854 | 0.9926 | 0.9890 | 136 |
| | | Stroma | 0.9932 | 1.0000 | 0.9966 | 146 |
| | | Tumour | 1.0000 | 0.9852 | 0.9925 | 135 |

| | | Accuracy | | | 0.9928 | 417 |
|---|---|---|---|---|---|---|
| | | Macro Avg | 0.9929 | 0.9926 | 0.9927 | 417 |
| | | Weighted Avg | 0.9929 | 0.9928 | 0.9928 | 417 |

4.3 Confusion Matrix Analysis

Figure 5 presents the confusion matrices, illustrating that the distilled student produces very few and clinically plausible errors. On Dataset-1, three classes—0_normal, 1_ulcerative_colitis, and 3_esophagitis—were perfectly classified (225/225 correct each). The only misclassifications occurred within inflammatory findings: two 2_polyps cases were predicted as 1_ulcerative_colitis, resulting in 223/225 correct for 2_polyps and an overall test accuracy of 0.9978. This pattern indicates that the decision boundaries are most distinct for normal tissue and esophagitis, with minor ambiguity arising between polyps and ulcerative colitis, a plausible overlap given the similarity of mucosal appearances in limited patches [47]. On Dataset-2, class separation was similarly strong. Stroma was fully recalled (146/146), the Other category was correctly identified in 135/136 cases (one misclassified as Stroma), and Tumour achieved 133/135 correct predictions, with the two misclassifications mapped to Other. Importantly, no non-tumour samples were predicted as Tumour, preserving Tumour precision at 1.0000 and minimizing false-positive escalation. The overall test accuracy was 0.9928. Collectively, the error patterns are asymmetric in a clinically favorable manner: misclassifications occur primarily between adjacent or histologically related categories, the model avoids overcalling malignancy, and stromal tissue in Dataset-2 as well as several classes in Dataset-1 are recognized perfectly. These results indicate well-calibrated decision boundaries rather than random confusion.

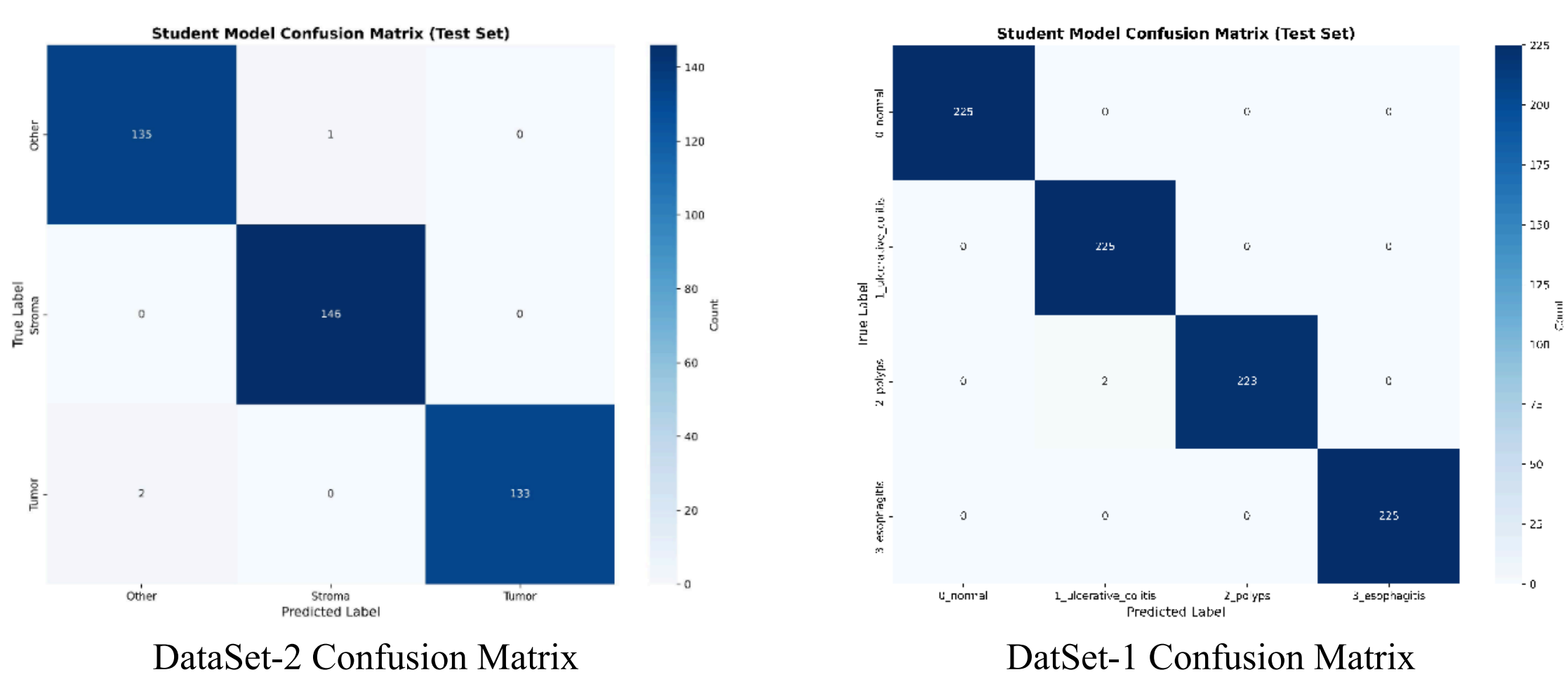

DataSet-2 Confusion Matrix

DatSet-1 Confusion Matrix

Fig 5: Confusion Matrix for both Datasets

4.2 Model Performance (Loss & Accuracy)

For Dataset-1, the student model exhibited a rapid ascent in training accuracy, exceeding 98% within the initial epochs and asymptotically reaching 99.9% by the conclusion of training [48]. Validation accuracy closely mirrored this trend, stabilizing at 99.7%, which aligns with the final reported validation and test accuracies of 99.67% and 99.78%, respectively. The loss curve followed an inverse trajectory, decreasing sharply from approximately 0.20 to below 0.02 before plateauing. This sustained convergence indicates an optimal training trajectory with negligible variance between splits, confirming robust generalization and resistance to overfitting.

A consistent learning pattern was observed for Dataset-2. The student achieved near-perfect training accuracy while maintaining a validation accuracy of approximately 99.3%, consistent with the final test accuracy of 99.28%. As illustrated in Fig. 6, training loss plummeted from 0.10 to approximately 0.01, while validation loss stabilized between 0.03 and 0.04 during the later epochs.

The tight alignment between training and validation performance across both datasets underscores the model's architectural stability. These results collectively demonstrate that the distilled student effectively inherits the teacher's discriminative power while leveraging knowledge distillation as a potent regularization mechanism for gastrointestinal image classification [49].

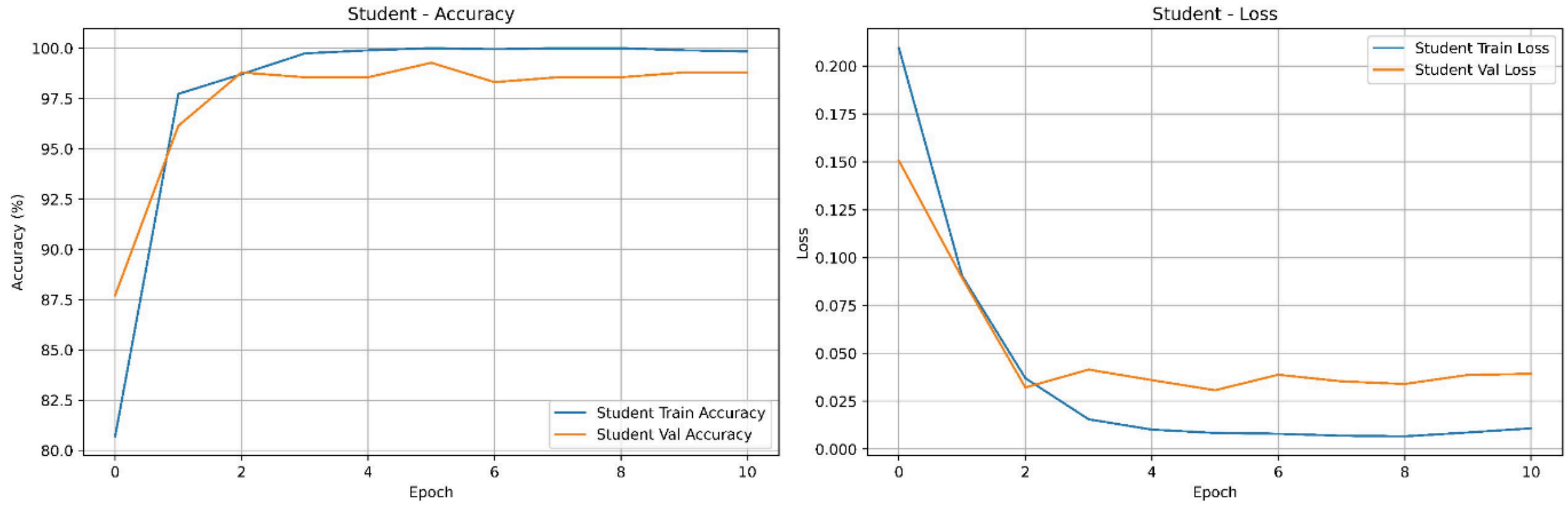

6(a) Dataset-2 Student Model Accuracy and Loss Graph

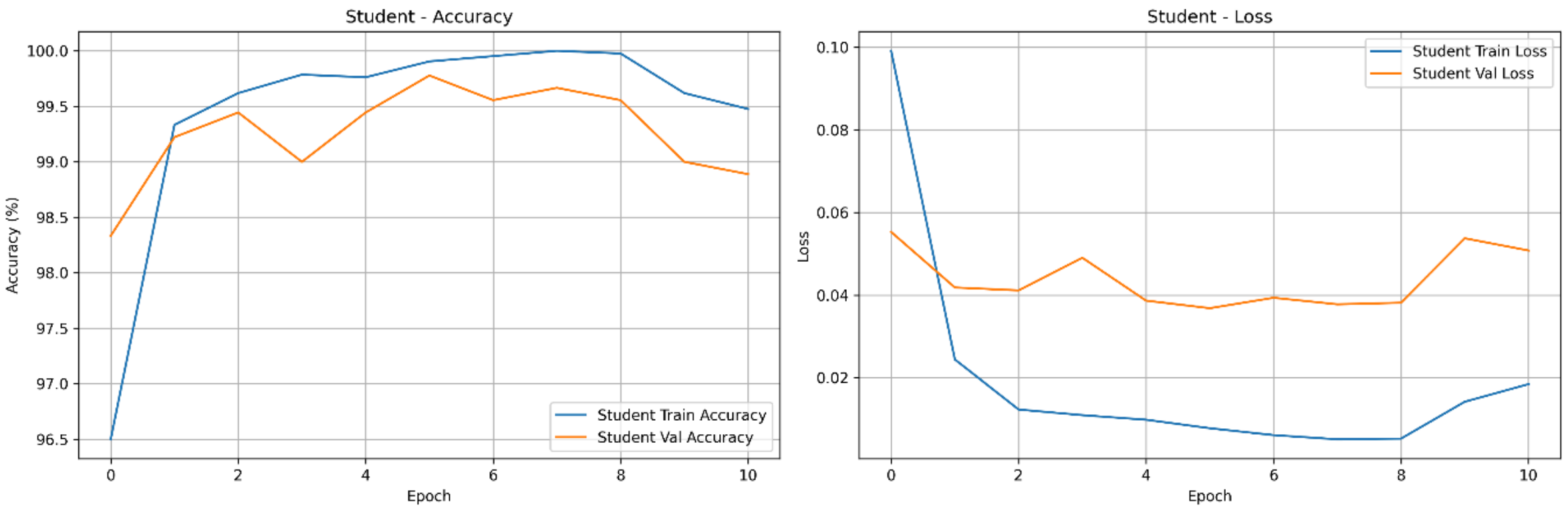

6(b) Dataset-1 Student Model Accuracy and Loss Graph

Fig 6. Training and Validation Accuracy and Loss Graph of the Student Models

## 4.3 ROC and AUC Analysis

The discriminative performance of the student model was further evaluated using Receiver Operating Characteristic (ROC) curve analysis across both datasets, as shown in Figure 7. The ROC curves for all classes are positioned near the upper-left corner, indicating low false-positive rates and high true-positive rates across decision thresholds.

For Dataset-1, the model demonstrated close to ideal class separability, with Area Under the Curve (AUC) values of 1.0000 for 0_normal, 2_polyps, and 3_esophagitis, and 0.9999 for 1_ulcerative_colitis. The macro-average AUC of 1.0000 confirms excellent discrimination among the four disease categories [50]. The close alignment of the ROC curves suggests that the model reliably distinguishes subtle variations in mucosal appearance across classes.

Similarly, on Dataset-2, the ROC curves indicate strong discriminative performance. AUC values were 0.9999 for Tumour, 0.9995 for Stroma, and 0.9980 for Other, resulting in a macro-average AUC of 0.9994. These results demonstrate that the student model achieves near-perfect class separation, particularly for the clinically critical Tumour category, where both high sensitivity and specificity are essential.

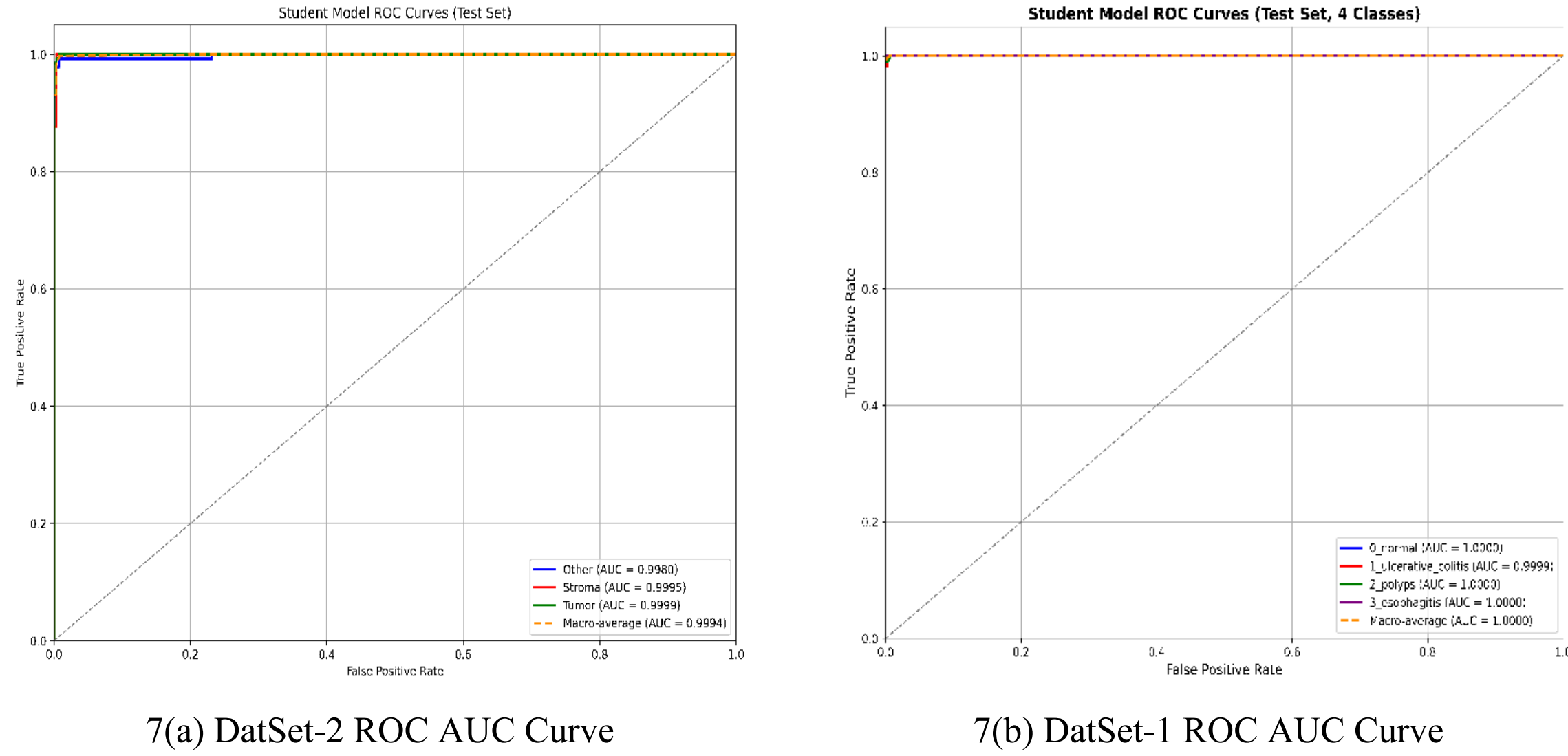

7(a) DatSet-2 ROC AUC Curve　　　　7(b) DatSet-1 ROC AUC Curve

Fig 7. ROC AUC CURVE for both Datasets

4.4 Qualitative Interpretability with Grad-CAM and Grad-CAM++

To assess whether the high quantitative performance of the proposed framework is supported by meaningful visual reasoning, qualitative interpretability analyses were conducted for both datasets using Gradient-weighted Class Activation Mapping (Grad-CAM) and Grad-CAM++ (Fig. 8). These visualization techniques produce heatmaps highlighting spatial regions most influential in the model's decisions, enabling evaluation of whether the network focuses on clinically and morphologically relevant structures [51].

For Dataset-2, which includes histopathological images categorized as Tumour, Stroma, and Other, Grad-CAM and Grad-CAM++ visualizations revealed well-localized, diagnostically relevant activation patterns. In Tumour images, the model primarily attended to dense malignant epithelial clusters and atypical nuclear regions, which are key pathological indicators of neoplastic tissue. For Stroma, highlighted regions corresponded to fibrous connective tissue and stromal cell distributions, while Other samples showed diffuse attention to non-malignant structures such as debris, lymphocytes, and normal epithelial areas. These results indicate that the model's feature extraction is guided by biologically and diagnostically meaningful cues rather than arbitrary patterns.

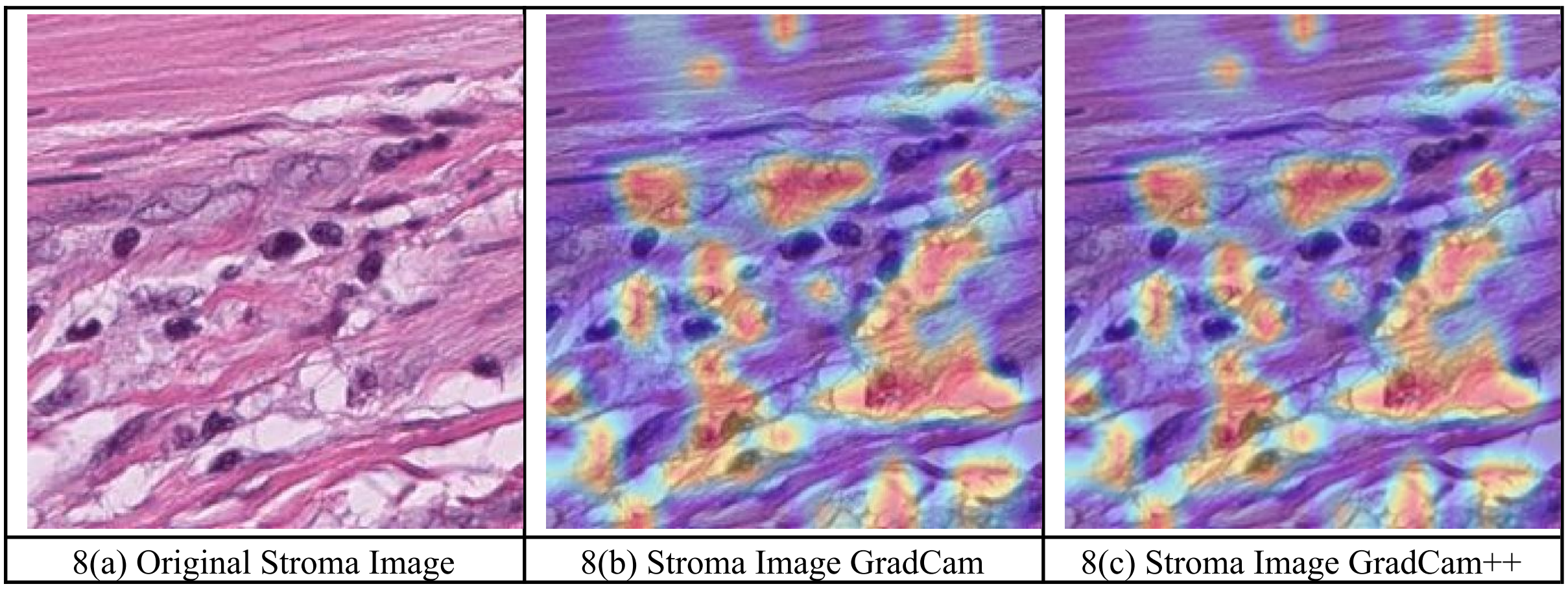

| 8(a) Original Stroma Image | 8(b) Stroma Image GradCam | 8(c) Stroma Image GradCam++ |
|---|---|---|

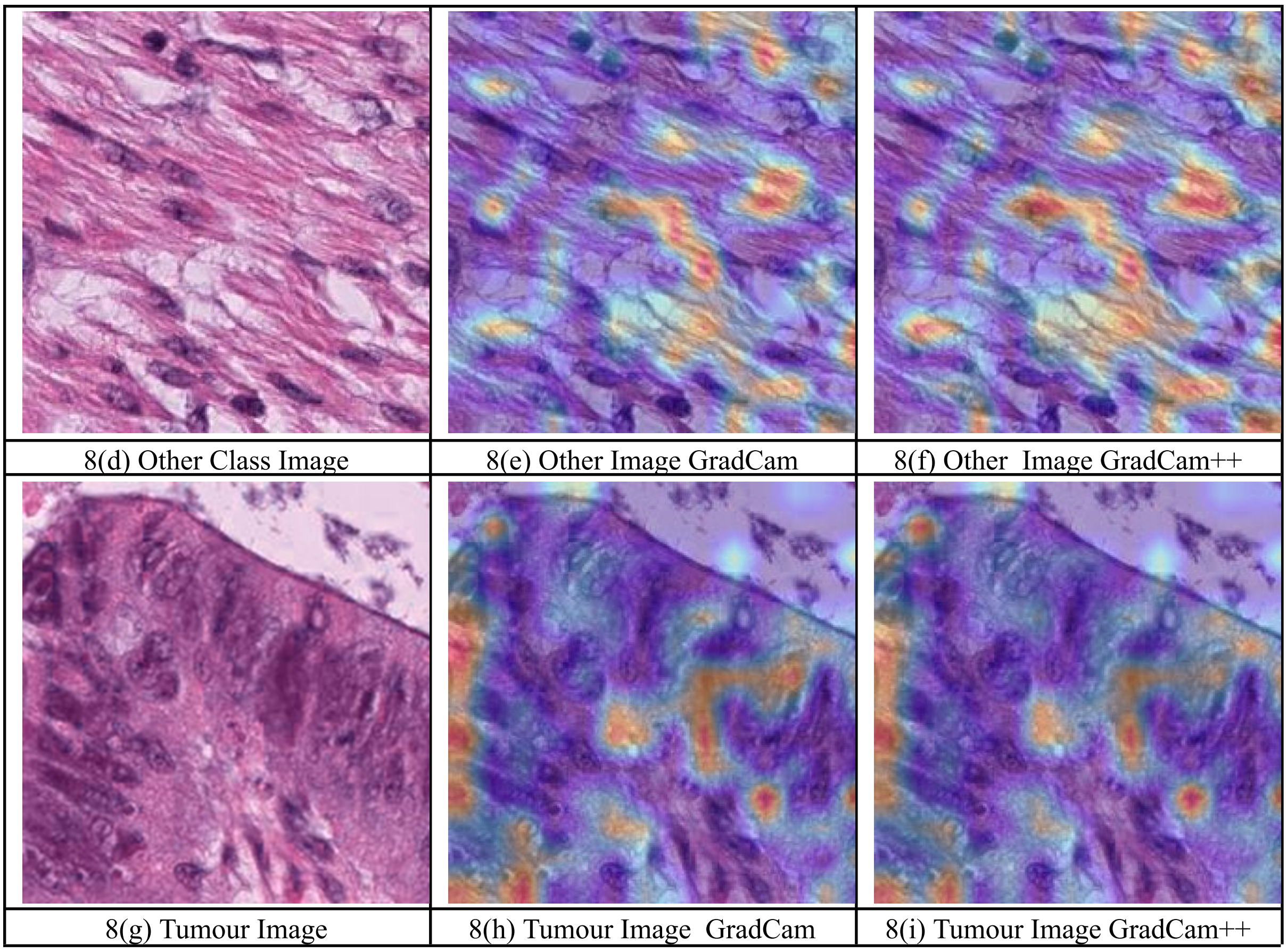

8(d) Other Class Image | 8(e) Other Image GradCam | 8(f) Other Image GradCam++

8(g) Tumour Image | 8(h) Tumour Image GradCam | 8(i) Tumour Image GradCam++

Fig 8. GradCam & GradCam++ for DataSet-2

For Dataset-1, which encompasses four gastrointestinal disease categories (normal, ulcerative colitis, polyps, and esophagitis), the activation maps revealed distinct, anatomically coherent attention patterns in Fig 9. For normal samples, the highlighted regions were uniformly distributed over the mucosal surface, consistent with the absence of pathological disruption. [52]

In ulcerative colitis, the Grad-CAM maps highlighted inflamed epithelial regions and crypt distortions, which are characteristic of mucosal inflammation. For polyps, the model focused on localized protruding tissue regions corresponding to adenomatous growths, while in esophagitis, activations were centered on epithelial disruptions and inflammatory foci. These consistent, pathology-specific attention patterns indicate that the student model's diagnostic reasoning aligns with clinically recognized visual markers rather than spurious background features.

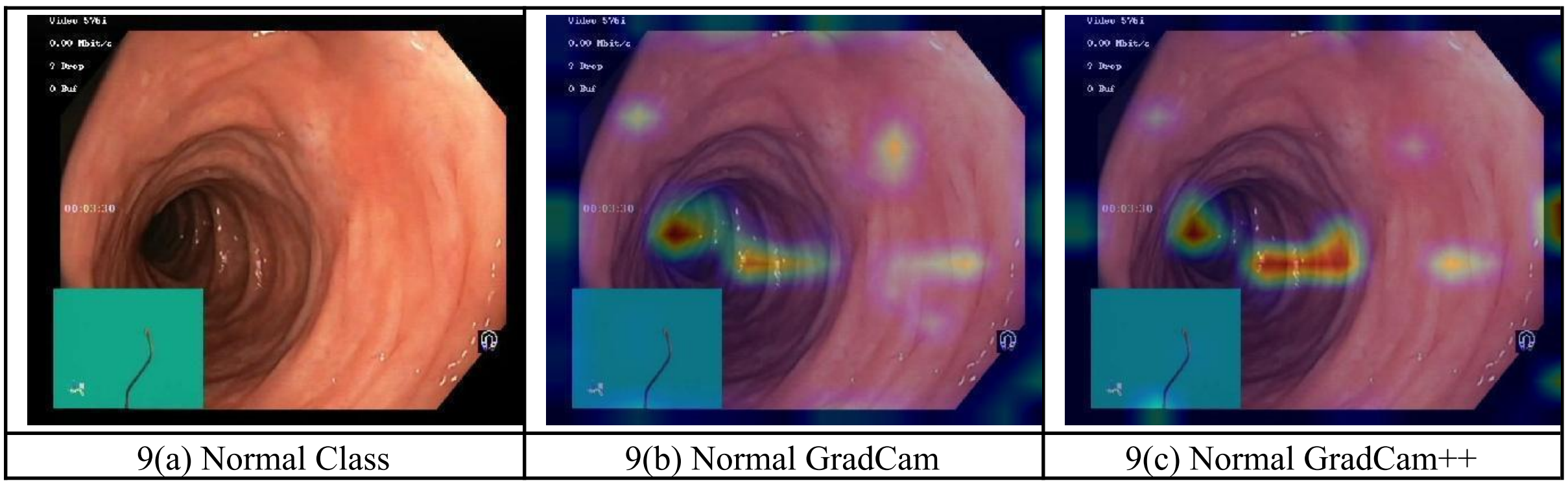

9(a) Normal Class | 9(b) Normal GradCam | 9(c) Normal GradCam++

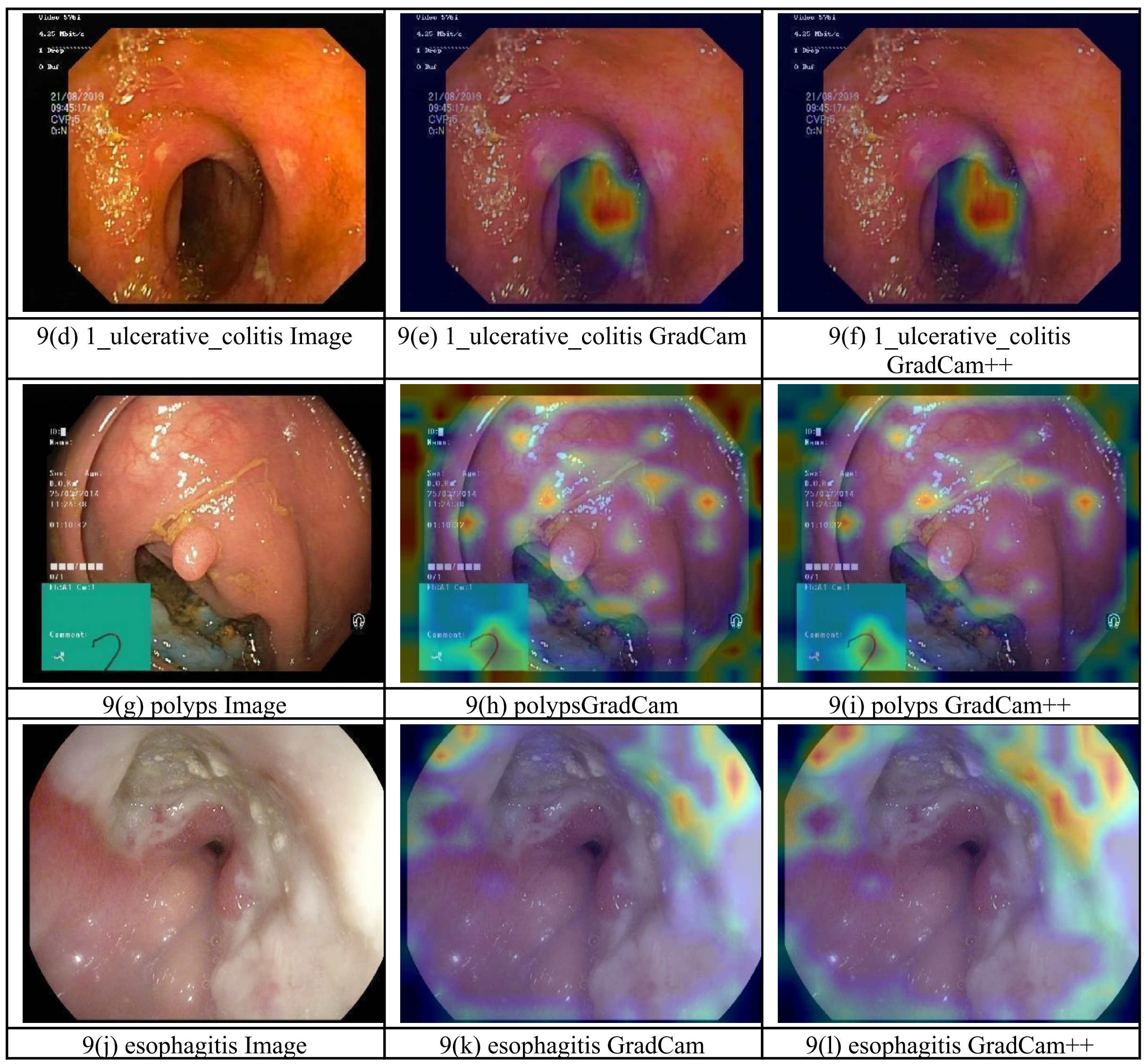

9(d) 1_ulcerative_colitis Image | 9(e) 1_ulcerative_colitis GradCam | 9(f) 1_ulcerative_colitis GradCam++

9(g) polyps Image | 9(h) polypsGradCam | 9(i) polyps GradCam++

9(j) esophagitis Image | 9(k) esophagitis GradCam | 9(l) esophagitis GradCam++

Fig 9. GradCam & GradCam++ for DataSet-1

4.5 Explainability Analysis Using LIME

To further enhance the interpretability of the proposed framework, Local Interpretable Model-Agnostic Explanations (LIME) was applied on Dataset-1 which is shown in Fig 8. Unlike Grad-CAM and Grad-CAM++, which provide class-discriminative visualizations at the feature-map level, LIME offers a complementary perspective by generating instance-level explanations that highlight the specific regions of each input image that most influenced the model's final prediction. [53] For each test image, LIME perturbed local regions (superpixels) and observed the corresponding changes in the model's output probabilities to approximate a locally interpretable linear model. The resulting visual overlays demonstrated that the student model relied on clinically relevant structures rather than irrelevant background cues. In *ulcerative colitis* samples, the highlighted regions corresponded to inflamed epithelial layers and distorted crypt patterns characteristic features of mucosal inflammation. For *polyps*, LIME consistently emphasized protruding glandular regions that aligned with pathological growths, while for *esophagitis*, the focus remained on epithelial disruptions and inflammatory lesions. In *normal* samples, the activation was distributed uniformly, indicating that the model recognized structural regularity as a key indicator of normalcy. These LIME-based visual explanations confirm that the model's high classification performance is rooted in medically interpretable reasoning rather than statistical artifacts. The correspondence between LIME saliency maps and pathologist-identified tissue regions supports the clinical relevance and interpretability of the distilled student model in gastrointestinal disease classification.

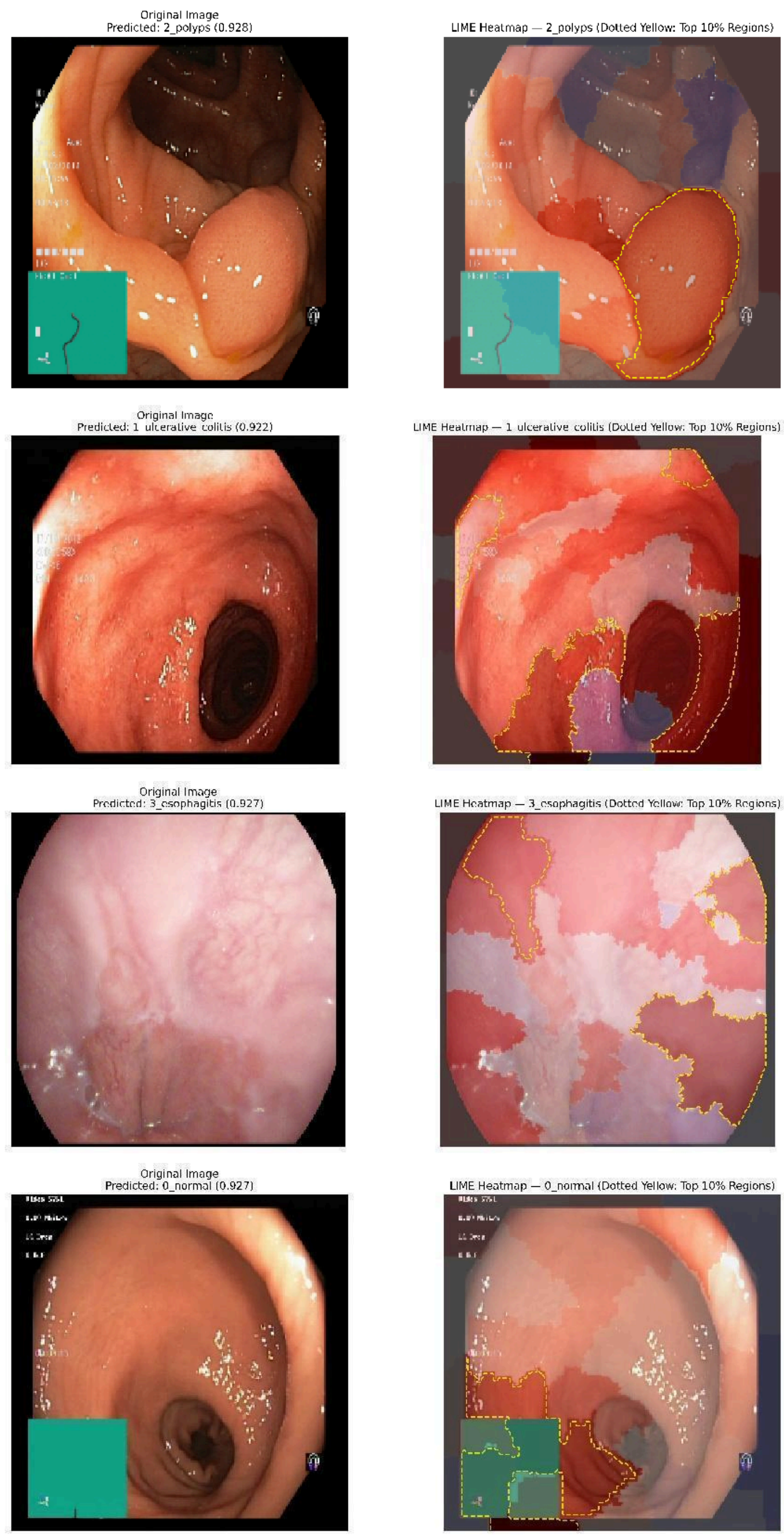

Fig 10. LIME visualization for DataSet-1

## 4.6 Explainability Analysis Using Score-CAM

To further assess the visual reasoning capability of the proposed model, Score-Weighted Class Activation Mapping (Score-CAM) was applied to both Dataset-1 and Dataset-2 shown in Fig 11. Score-CAM extends conventional gradient-based interpretability techniques such as Grad-CAM by using the model's own class confidence scores, rather than backpropagated gradients, to weight activation maps. This gradient-free approach reduces noise and produces more stable and visually precise heatmaps, offering a clearer understanding of the discriminative regions influencing each prediction. For Dataset-1, which includes normal, ulcerative colitis, polyps, and esophagitis classes, the Score-CAM visualisations revealed strong correspondence between the highlighted image regions and known pathological indicators. [54,55] In ulcerative colitis and esophagitis, the model primarily attended to epithelial disruptions, inflammatory zones, and crypt distortions, consistent with clinical findings. For polyps, the heatmaps concentrated on local glandular protrusions and tissue irregularities associated with neoplastic growth. Meanwhile, standard samples showed evenly distributed activations over the mucosal surface, reflecting the absence of structural abnormalities. These observations confirm that the model's feature extraction aligns with expert-recognised diagnostic regions.

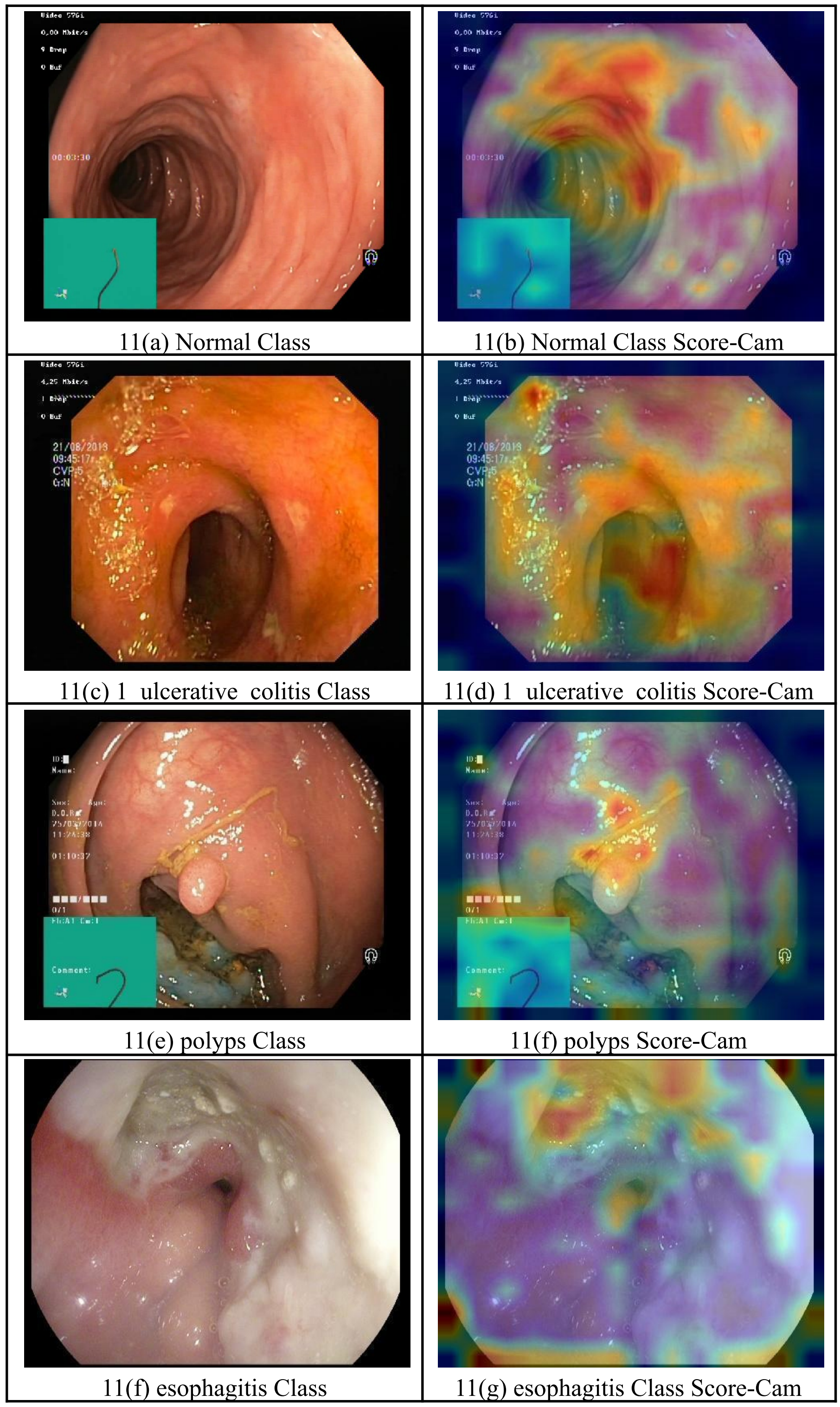

Fig 11. Score Cam Visualisation on Dataset-1

For Dataset-2, comprising Tumour, Stroma, and other tissue classes, Score-CAM maps similarly demonstrated anatomically meaningful attention. In Tumour samples, activation hotspots corresponded to malignant epithelial clusters and dense nuclei hallmarks of neoplastic pathology. Stroma visualizations emphasized fibrous connective structures, while other images showed diffuse focus over non-malignant tissue areas, including debris and lymphocytic zones. These class-specific activation patterns highlight the model's ability to discriminate between cancerous and non-cancerous regions with high spatial precision shown in Fig 12.

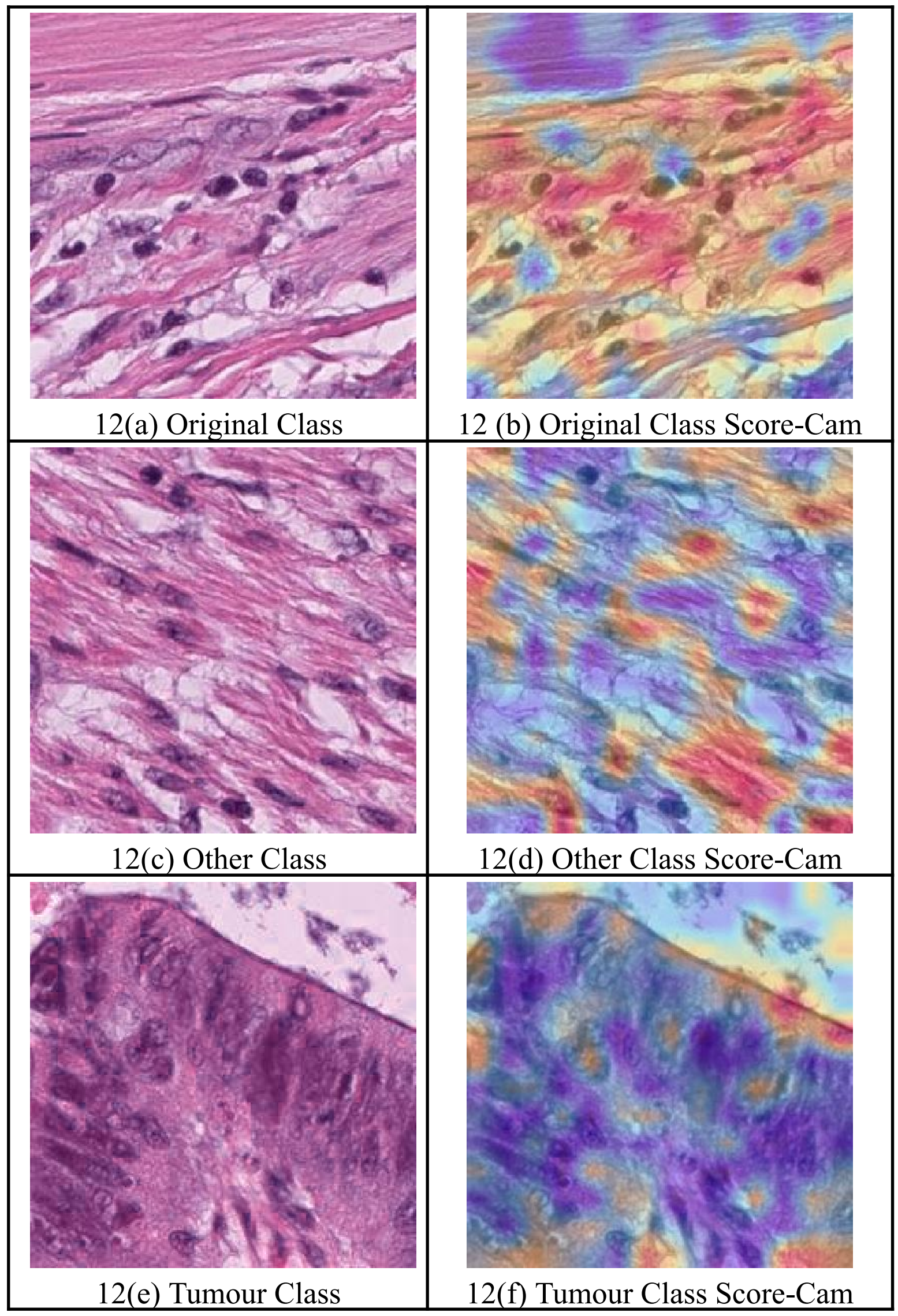

Fig 12. Score-Cam of DataSet-2

Table 8. Comparative Analysis of the Results with Relevant Research

| Author / Year | Dataset Used | Model / Technique | Accuracy / AUC |
|---|---|---|---|
| Le et al., 2025 [7] | Public CRC histopathology dataset | CNN-based hierarchical classifier | 98.9 % / 0.995 |
| Ömeroğlu, 2024 [8] | NCT-CRC-HE-100K | Supervised contrastive + ResNet backbone | 99.1 % |
| Fadafen & Rezaee, 2023 [9] | CRC-TP dataset | Ensemble hybrid CNN + transformer | 99.2 % |
| El Amine et al., 2023 [10] | TCGA-COAD, LUAD | ViT + GAN + KD framework | 98.5 % |

| Ke et al., 2025 [11] | CRC-100K, BreakHis | CNN ensemble (ResNet, DenseNet) | 99.3 % |
|---|---|---|---|
| Mirza et al., 2025 [12] | Private WCE dataset | Fused deep + shallow CNNs | 98.6 % |
| Bordbar et al., 2023[13] | Kvasir-Capsule | 3D CNN (ResNet-50 3D) | 99.0 % |
| Shukla & Jayaraman, 2025 [14] | Kvasir-V2 | CNN + Grad-CAM/LIME | 98.2 % |
| Attallah et al., 2025 [15] | Kvasir, HyperKvasir | EndoNet (multiscale CNN + attention) | 99.5 % |
| Tan et al., 2024 [16] | Kvasir-Capsule | EndoOOD (ViT + uncertainty estimation) | 97.8 % |
| Ömeroğlu, 2021 [17] | CRC-100K | Transfer learning with EfficientNet | 98.7 % |
| Alotaibi et al., 2024 [18] | Private CRC slides | HIELCC-EDl (Wiener filter + swarm optimizer) | 99.6 % |
| Li et al., 2024 [19] | 1,000 CRC cases (multistain) | MSDLM (CNN + transformer) | 99.4 % |
| Huang et al., 2025 [20] | Clinical Stage II CRC cohort | CT-Path fusion model (DeepSurv + CNN) | C-index 0.83 |
| Subramanian et al., 2023 [21] | 256 patient rectal tissue samples | csPWS (optical spectroscopy + DL) | AUC 0.96 |
| Our Proposed Approach | Kaggle WCE dataset, Mendeley Colorectal Histopathology dataset | KD-TinyViT-LoRA | 99.78%, 99.28% |

## 5. Conclusion and Future Works

This study introduced a hybrid dual-stream deep learning framework founded on knowledge distillation between a high-capacity teacher model integrating a Swin Transformer (Small) and a Vision Transformer (ViT-16 Small), and a lightweight Tiny-ViT (5 M parameters) student model for colorectal and gastrointestinal (GI) disease classification. Leveraging two curated and balanced histopathological and endoscopic datasets, the proposed approach achieved close to ideal performance across all evaluation metrics, demonstrating exceptional accuracy, generalization, and clinical interpretability.

The teacher–student distillation mechanism allowed the Tiny-ViT model to inherit the global contextual reasoning of the Swin Transformer and the fine-grained feature extraction of the ViT encoder, effectively combining semantic richness with computational efficiency. This design reduced computational overhead while maintaining high diagnostic precision, supporting potential deployment in real-time and resource-constrained clinical settings. Across both datasets, the student model achieved average accuracy exceeding 99.7%, with minimal variation between training, validation, and test splits, indicating robust and stable performance.

Qualitative interpretability analyses using Grad-CAM, Grad-CAM++, LIME, and Score-CAM confirmed that the model's decisions were grounded in anatomically and pathologically meaningful regions, supporting its reliability as a clinically interpretable computer-aided diagnostic (CAD) system.

Despite these promising results, several limitations remain. The datasets, although balanced and curated, are limited in size and diversity compared to large-scale clinical repositories, which may affect generalization to unseen imaging conditions. Additionally, the current framework focuses on image-level classification and does not exploit temporal continuity present in sequential endoscopic or MRI data.

Despite these promising results, several challenges remain. The datasets employed, though well-balanced and curated, are limited in size and diversity relative to large-scale real-world clinical repositories, which may constrain the model's adaptability to unseen imaging conditions. Furthermore, the present framework focuses primarily on image-level classification, without exploiting the temporal continuity inherent in sequential endoscopic or MRI imaging. Expanding dataset diversity through multi-center and cross-device collections to enhance generalization and mitigate domain bias. Temporal sequence modeling using video-based transformers or recurrent attention mechanisms to capture dynamic lesion evolution in endoscopic recordings. Integrating multimodal information, such as patient metadata and clinical reports, to develop a holistic diagnostic pipeline. Model compression and edge deployment, optimizing Tiny-ViT for on-device inference and real-time clinical use. Uncertainty-aware explainable AI, incorporating Bayesian or probabilistic layers to quantify prediction confidence and improve interpretability.

In conclusion, the Swin + ViT–to–Tiny-ViT knowledge-distillation framework establishes a robust, interpretable, and computationally efficient foundation for AI-assisted gastrointestinal disease diagnosis. By extending toward scalable, multimodal, and explainable implementations, this work provides a step toward intelligent diagnostic systems that can support clinicians and improve patient outcomes.